%Paper: cond-mat/9304020
%From: privman@craft.camp.clarkson.edu (Prof. V. Privman)
%Date: Fri, 9 Apr 93 20:14:02 -0400

% The paper is in plain TeX  and it can be
% TeXed directly as provided. No input files,
% etc., are needed.

\def\NI{\noindent}

\magnification=\magstep 1
\overfullrule=0pt
\hfuzz=16pt
\voffset=0.0 true in
\vsize=8.8 true in

\baselineskip 12pt
\parskip 2pt
\hoffset=0.1 true in
\hsize=6.3 true in
\nopagenumbers
\pageno=1
\footline={\hfil -- {\folio} -- \hfil}

\centerline{\bf Dissociative Adsorption: A Solvable Model}

\vskip 0.4in

\centerline{\bf Vladimir~Privman}

\vskip 0.2in

\centerline{Department of Physics, Clarkson
University, Potsdam, New York 13699--5820, USA}

\vskip 0.4in

\NI\hang {\bf Abstract:}\ \ A model of ``hot''-dimer deposition in one
dimension, introduced by Pereyra and
Albano, is modified to have an unbounded dissociation range.
The resulting dynamical equations are solved exactly. A related
$k$-mer dissociation model is also introduced and its solution obtained
as a quadrature.

\

\NI {\bf PACS numbers:}$\;$  68.10.Jy {\it and\/} 82.65.-i

\vskip 0.4in

Recent experimental studies [1] reported deposition processes in
which prior to adhesion the arriving diatomic molecules break up into
single-atom fragments which dissipate
the excess energy by ``flying apart'' a large
distance. A simplified one-dimensional (1D) model of these processes was
introduced in [2]. The restriction to 1D was motivated by the
possibility of extensive numerical Monte Carlo studies, and in
fact, more realistic two-dimensional-substrate Monte Carlo modeling was
alluded to in [2]. It is usually instructive to obtain solvable
variants of 1D models. Indeed, exact calculations
provide insight into more realistic higher-dimensional systems
and yield test models for approximation schemes, in various reaction and
adsorption processes reviewed, e.g., in [3-6].

In this work, we introduce a model with infinite dissociation range and
with random initial conditions, which can be solved exactly for dimers
as well as generalized to $k$-mer deposition with dissociation. The
latter generalization allows discussion of the approach to the continuum
limit [7-8].

In the deposition of ``hot'' dimers [2], one
assumes that each successful deposition
attempt in followed up by the dissociation
of the dimer. The two monomer fragments
move apart up to a certain
maximal distance: they either stop on their own or run
into existing clusters. In order to eliminate the upper
bound on the allowed dissociation distance
while still have the resulting model not
sensitive to the finite-size end-effects, one has to consider initial states in
which obstacles to the motion of the fragments after deposition are present
with
reasonable density. Thus we assume that
initially the lattice is randomly covered
with some finite monomer density $\rho$, measured per lattice site.

Each dimer deposition event will be followed by instantaneous fragmentation and
attachment of one monomer
fragment at each end of the empty gap in which the deposition
attempt has succeeded. Thus the gap size will be decreased by 2. For $k$-mer
deposition, with $k=3,4,5,\ldots$, the way in which
the $k$-mer breaks down into two
fragments is irrelevant as far as the effect of the deposition event on the gap
length
is concerned. Indeed, the gap will be shortened by $k$ sites no matter in what
proportion were the original $k$ monomers placed at each of its ends. Monomer
deposition, $k=1$, can be also treated, with obvious modifications.

The exact solvability of the resulting model is then nearly obvious. The gap
sizes
evolve independently, unlike in models with finite
dissociation range [2] and in related models [9] where the relaxation is by
diffusion. The fact that added deposition processes confined to the gap ends do
not
impede exact solvability has also been noted in random sequential adsorption
with
``ballistic relaxation'' [10].

It is convenient to consider a lattice of $N$ sites, although we will disregard
any
end effects (i.e., the limit $N \to \infty$ will be always assumed). Let
$P_m(t)N$
denote the number of gaps exactly $m$ empty sites long, at time $t$. The
initial random
distribution corresponds to

$$ P_m(0)=\rho^2 (1-\rho)^m \; , \eqno(1) $$

\NI where $m=0,1,2,\ldots $. The coverage fraction, $\theta (t)$, is initially
$\theta(0)=\rho$.

We assume that the deposition attempts are random and uncorrelated at various
lattice
sites, with the rate $R$ per site. Note that for an $N$-site lattice there are
$N$
distinct $k$-mer deposition locations (if we disregard end effects). The rate
$R$ is
defined per each lattice location, and it will be conveniently absorbed in the
definition of the dimensionless time variable,

$$ \tau = Rt \; . \eqno(2) $$

\NI Of course, only deposition attempts for which the $k$-mer fully fits in an
empty
gap are successful. Other attempts are rejected.

Since the fragments of the dissociation are transported to the ends of the
gaps, the
time evolution of the gap numbers satisfies a simple set of relations,

$$ {dP_m \over d\tau} = -(m-k+1)P_m +(m+1)P_{m+k} \qquad {\rm for} \qquad m
\geq k-1 \;
, \eqno(3) $$

$$ {dP_m \over d\tau} = (m+1)P_{m+k} \qquad {\rm for} \qquad 1 \leq m \leq k-1
\; .
\eqno(4) $$

\NI The relations (3) can be solved in closed form by standard methods such as
generating functions or other techniques developed for random sequential
adsorption
[5-6]. We only quote the result,

$$ P_{m \geq k-1} ( \tau ) = { \rho^2 (1- \rho )^m e^{-(m-k+1) \tau } \over
\left[ 1 - (1- \rho )^k \left(1- e^{-k \tau } \right) \right]^{(m+1)/k} } \;\;
{}.
\eqno(5) $$

The coverage can be obtained by summing up the monomer deposition
rates in the $m \geq k $ gaps,

$$ { d \theta \over d \tau } = k \sum\limits_{m=k}^\infty (m-k+1) P_m \; .
\eqno(6) $$

\NI This relation leads to a quadrature which could not be
evaluated in closed form for general $k$; see (12) below. Similarly evaluation
of the
gap numbers for $1 \leq m \leq k-1$ requires integration (for $k>2$); see (4).
Thus, an
alternative expression for the coverage via the fraction of unoccupied sites,

$$ 1- \theta = \sum\limits_{m=1}^\infty m P_m \; , \eqno(7) $$

\NI still involves quadratures for general $k$.

We are particularly
interested in the cases $k=1,2$ and $k \to \infty$. Fortunately, for all these
values
of $k$ further progress is possible. We report the results for each case in
turn.
First, consider the monomer deposition process which is expected to proceed to
full
coverage at large times. Indeed, the exact expression is simply

$$ \theta_{k=1} (\tau ) = 1- (1-\rho ) e^{-\tau } \; . \eqno(8) $$

Calculation of the coverage for the dimer deposition, $k=2$, is also
straightforward. Indeed, all the terms needed in (7) are given by the solution
(5)
(which is no longer the case for $k>2$). The resulting expression,

$$ \theta_{k=2} (\tau )  = 1- {1- \rho \over (2- \rho )^2 }
\left[ (2- \rho ) \rho + 2 (1- \rho )^2 e^{-2\tau }
 + 2 (1- \rho ) e^{-\tau } \sqrt{ 1 - (1- \rho )^2
 \left(1- e^{-2 \tau } \right) } \right] \;\; , \eqno(9) $$

\NI was also checked by calculating (6).

Since each gap shrinks independently of all other gaps in this model, the
evolution of
the gap size distribution could also be studied in terms of the probability
distribution of stochastic gap-size variables $m(t)$ decreasing in
steps of $k$ at the rates proportional to $m(t)-k+1$ for $m(t) \geq k$, and
zero for
$0 \leq m(t)<k$. The value $m(0)$ can be considered fixed for computational
purposes,
and the final expressions averaged over the initial $m$-value distribution (1).
For
finite times this approach does not seem to yield new useful results. However,
for time
$t=\infty$ all $m(0)$-size gaps will be reduced mod$(k)$ to the values
$m(\infty)=0,1,\ldots,k-1$. Therefore the $t=\infty$ ``jamming'' coverage can
be
calculated as

$$  \theta (\infty) =1- \sum\limits_{m=1}^{k-1} m \sum\limits_{j=0}^\infty
P_{m+jk}(0) = \rho \left[1+{k(1-\rho)^k \over 1-(1-\rho)^k  } \right] \; .
\eqno(10) $$

The $\tau=\infty$ values for $k=1,2$ are consistent with (8) and (9). Note that
the
limiting coverage for $k=2$ is less than 1, and it depends on the initial
density
$\rho$ via

$$ \theta_{k=2} (\infty) = {1+(1-\rho)^2 \over 2-\rho} \; . \eqno(11) $$

\NI An interesting property of the jamming coverages (10) is that they are
nonmonotonic
functions of $\rho$ for $0<\rho \leq 1$. The rate of approach to the jamming
limit is
$\sim e^{-\tau}$.

We now turn to the general-$k$ relation (6) which can be reduced to the form

$$ {d \theta \over d \tau } = { k \rho^2 (1-\rho )^k e^{-\tau } \over
\left[ 1-(1-\rho )^k \left(1-e^{-k\tau } \right) \right]^{(k-1)/k} }
\left\{ \left[ 1-(1-\rho )^k \left(1-e^{-k\tau } \right) \right]^{1/k}
-(1- \rho ) e^{-\tau } \right\}^{-2} \;\; . \eqno(12) $$

\NI In order to formulate the continuum limit as $k \to \infty$, similar to the
procedure described for random sequential adsorption in [7-8], we define the 1D
lattice with spacing decreasing as $\sim {1 \over k}$. The depositing
$k$-mer objects then have fixed length. In order to keep their deposition
attempt
rate fixed per unit length which for $k \to \infty$ accommodates O$(k)$ lattice
sites, we define the deposition rate per site, $R$, to be proportional to
$1/k$.
Then the time variable finite in the continuum limit is [7-8]

$$ T=k \tau \; . \eqno(13) $$

The present problem, however, has a new interesting aspect in the continuum
limit,
not discussed in connection with random sequential adsorption studied in [7-8].
Indeed
the initial conditions of placing monomers randomly with density $\rho$ per
site,
correspond to the average length of empty gaps equal $1/ \rho$ lattice spacings
as can
be easily verified from (1). Thus for $\rho$ fixed the deposition process is
blocked in
the limit $k \to \infty$, i.e., the coverage remains $\rho$, in fact, for times
$\tau
< {\rm O}(k)$. This property is also related to the fact that the jamming
coverages
(10) approach asymptotically the curve $\theta (\infty) = \rho$ when the $k \to
\infty$
limit is taken for fixed $\rho>0$. To have a nontrivial limit, one has to take
$\rho$
of order $1/k$. This corresponds to vanishing initial coverage but to finite
density
of zero-length ``obstacles'' that stop the depositing object fragments upon
dissociation. We take

$$ \rho = {1 \over Mk} \; , \eqno(14) $$

\NI corresponding to the average distance of $Mk$ lattice spacings between
these
``obstacles'' which serve as seeds for the occupied-area ``clusters'' formed at
finite times $T$.

The limiting form of (12) as $k \to \infty$ is

$$ {d \theta_{k=\infty} \over dT} = {1 \over M^2 \left( e^{1/M} -1 +e^{-T}
\right)
\left[ T + \ln \left( e^{1/M} -1 +e^{-T} \right) \right]^2 } \;\; , \eqno(15)
$$

\NI while the jamming coverage (10) reduces to

$$ \theta_{k=\infty} (\infty) = { 1\over M \left( e^{1/M} -1  \right) } \; .
\eqno(16) $$

\NI Thus the deposition process develops the power-law tail,

$$ \theta_{k=\infty} (\infty) - \theta_{k=\infty} (T) \simeq
{ 1\over M^2 \left( e^{1/M} -1  \right) T } \qquad {\rm for} \qquad T \gg 1  \;
{}.
\eqno(17) $$

\NI This power-law behavior is similar to the ordinary random
sequential adsorption [7-8]: the tail is due to gaps arbitrary close in
size to that of the depositing objects. These gaps are reached with small
probability
by uniformly distributed deposition attempts. For short times
the continuum-limit coverage behaves according to

$$\theta_{k=\infty} (T) \simeq e^{1/M} T  \qquad {\rm for} \qquad T \ll 1  \; .
\eqno(18) $$

Finally, we point out that the $\rho \to 0$ limit of the coverage exists and is
well
defined for all $k=1,2,3,\ldots,\infty$. The result corresponds to uncorrelated
deposition of $k$-mers,

$$ \theta_{\rho \to 0}(\tau) = 1-e^{-k\tau} = 1-e^{-T} \; . \eqno(19) $$

\NI This expression follows by taking the $\rho \to 0$ limit in (8), (9), (12),
or the
$M \to \infty$ limit in (15).

In summary, we introduced a variant of the 1D dissociative adsorption process
which
allows exact solution for dimers as well as derivation of various other results
notably the form of the continuum behavior for limiting off-lattice deposition.
The
results for the time dependence of the coverage generally resemble those for
random
sequential adsorption [5-8]. A new feature is the form of the initial
conditions
required for the existence of the continuum limit.

\

\centerline{\bf REFERENCES}

{\frenchspacing

\item{[1]} H. Brune, J. Wintterlin, R.J. Behm and G. Ertl,
Phys. Rev. Lett. {\bf 68}, 624 (1992).

\item{[2]} V.D. Pereyra and E.V. Albano,
{\sl Random Sequential Adsorption of ``Hot'' Dimers on One
Dimension}, J. Phys. A, in print (1993).

\item{[3]} T. Liggett, {\sl Interacting Particle Systems\/}
(Springer-Verlag, New York, 1985).

\item{[4]} V. Kuzovkov and E. Kotomin, Rep. Prog. Phys.
{\bf 51}, 1479 (1988).

\item{[5]} M.C. Bartelt and V. Privman, Int. J. Mod. Phys.
B{\bf 5}, 2883 (1991).

\item{[6]} J.W. Evans, {\sl Random and Cooperative Sequential Adsorption},
Rev. Mod. Phys., in print (1993).

\item{[7]} J.J. Gonzalez, P.C. Hemmer  and J.S. H{\o}ye, Chem. Phys.
{\bf 3}, 228 (1974).

\item{[8]} V. Privman, J.--S. Wang and P. Nielaba, Phys. Rev.
B{\bf 43}, 3366 (1991).

\item{[9]} M.C. Bartelt and J.W. Evans, Phys. Rev.
B{\bf 46}, 12675 (1992).

\item{[10]} J. Talbot and S.M. Ricci, Phys. Rev. Lett. {\bf 68}, 958 (1992).

}

\bye